\documentclass[10pt, conference, compsocconf]{IEEEtran}
\IEEEoverridecommandlockouts
\usepackage{graphicx}
\graphicspath{{figures/}}
\usepackage{float}
\usepackage{multirow}
\usepackage{booktabs}
\usepackage{array}
\usepackage{caption}
\usepackage{comment}
\let\labelindent\relax
\usepackage{enumitem}
\usepackage{arydshln}
\usepackage{amsmath}
\usepackage{amsfonts}
\usepackage{nicefrac}
\usepackage{xcolor}
\usepackage{fdsymbol}
\usepackage{wrapfig}
\usepackage{bbm}
\usepackage{hyperref}

\newcommand{\tabincell}[2]{\begin{tabular}{@{}#1@{}}#2\end{tabular}} 

\hypersetup{
    colorlinks=true,
    linkcolor=blue,
    filecolor=magenta,      
    urlcolor=black,
    citecolor=black,
    pdfpagemode=FullScreen,
}

\newcommand{\hr}[1]{\textcolor{black}{#1}}
\setlist[itemize]{leftmargin=3mm}

\begin{document}
%
\title{GCoD: Graph Convolutional Network Acceleration via Dedicated \\ Algorithm and Accelerator Co-Design
\vspace{-0.2em}}




%

\author{
\IEEEauthorblockN{Haoran You$^*$\IEEEauthorrefmark{2},
Tong Geng$^*$\IEEEauthorrefmark{3},
Yongan Zhang\IEEEauthorrefmark{2}, 
Ang Li\IEEEauthorrefmark{3} and
Yingyan Lin\IEEEauthorrefmark{2}\thanks{* denotes equal contribution}\\}
\IEEEauthorblockA{\IEEEauthorrefmark{2}Rice University, Houston, TX\\}
\IEEEauthorblockA{\IEEEauthorrefmark{3}Pacific Northwest National Laboratory, Richland, WA\\}
\{haoran.you, yz87, yingyan.lin\}@rice.edu, \{tong.geng, ang.li\}@pnnl.gov
}


\maketitle

\begin{abstract}
Graph Convolutional Networks (GCNs) have emerged as the state-of-the-art graph learning model. However, it can be notoriously challenging to inference GCNs over large graph datasets, limiting their application to large real-world graphs and hindering the exploration of deeper and more sophisticated GCN graphs. This is because real-world graphs can be extremely large and sparse. Furthermore, the node degree of GCNs tends to follow the power-law distribution and therefore have highly irregular adjacency matrices, resulting in prohibitive inefficiencies in both data processing and movement and thus substantially limiting the achievable GCN acceleration efficiency. 
To this end, this paper proposes a \textbf{G}CN algorithm and accelerator \textbf{Co}-\textbf{D}esign framework dubbed GCoD which can largely alleviate the aforementioned GCN irregularity and boost GCNs' inference efficiency. Specifically, \underline{on the algorithm level}, 
GCoD integrates a split and conquer GCN training strategy that polarizes the graphs to be either denser or sparser in local neighborhoods without compromising the model accuracy, resulting in graph adjacency matrices that (mostly) have merely two levels of workload and enjoys largely enhanced regularity and thus ease of acceleration.
\underline{On the hardware level}, we further develop a dedicated two-pronged accelerator with a separated engine to process each of the aforementioned denser and sparser workloads, further boosting the overall utilization and acceleration efficiency.
Extensive experiments and ablation studies validate that our GCoD consistently reduces the number of off-chip accesses, leading to speedups 15286$\times$, 294$\times$, 7.8$\times$, and 2.5$\times$ as compared to CPUs, GPUs, and prior-art GCN accelerators including HyGCN and AWB-GCN, respectively, while maintaining or even improving the task accuracy.
Additionally, we visualize GCoD trained graph adjacency matrices for a better understanding of its advantages.
Codes are available at \url{https://github.com/RICE-EIC/GCoD}.

\end{abstract}

\begin{IEEEkeywords}
GCNs; algorithm and accelerator co-design;

\end{IEEEkeywords}

%
\IEEEpeerreviewmaketitle

\vspace{-0.3em}
\section{Introduction}
\vspace{-0.1em}

The recent breakthrough achieved by deep learning has motivated growing demands for deep learning powered intelligence in many daily life devices featuring contrained resources and a small form factor \cite{8050797,10.1145/3210240.3210337}. 
For example, 
we have recently witnessed the tremendously increased excitement towards Graph Convolutional Networks (GCNs), which has achieved state-of-the-art (SOTA) performance for graph-based learning tasks. 
The superior performance of GCNs largely benefits from GCNs' irregular and unrestricted neighborhood connections via two primary execution phases: aggregation and combination, where the former maintains most graph processing behaviors and the latter acts more like neural networks. Specifically, during the aggregation phase, for each node in a graph, GCNs first aggregate all its neighbor nodes' features, which heavily relies on the graph structure that is inherently random and sparse; during the combination phase, GCNs transform the aggregated features through (hierarchical) feed-forward propagation to update the feature of the given node. 
In parallel, recent breakthroughs of GCNs have ignited an explosive interest in investigating GCNs for numerous real-world applications, including accurate advertisement in E-commerce \cite{yang2019aligraph} and electric grid cascading failure analysis \cite{liu2020guiding}.
Many of them impose stringent latency/throughput constraints, e.g., real-time decision-making.
The promising performance and the potential exciting applications of GCNs come with prohibitive challenges that limit their applications to large real-world graphs and hinder the exploration of deeper and more sophisticated GCN graphs. \underline{First},
graphs (or graph data), especially real-world ones, are often extraordinarily large and irregular as exacerbated by their intertwined complex neighbor connections, e.g., there are a total of 232,965 nodes in the Reddit graph with each node having about 50 neighbors \cite{KKMMN2016}. One outcome is that GCNs tend to follow the power-law distribution and therefore have highly irregular adjacency matrices resulting in prohibitive inefficiencies in both data processing and movement which substantially limits the achievable GCN acceleration efficiency. \underline{Second}, the dimension of GCNs' node feature vectors can be very high, e.g., each node in the Citeseer graph has 3703 features, which can lead to paramount processing costs in the combination phrase. 
As an illustrative example, a 2-layer GCN requires 19G FLOPs (FLOPs: floating-point operations) to process the Reddit graph \cite{tailor2020degree}, resulting in a latency of 2.94E5 milliseconds when executed on an Intel Xeon E5-2680 CPU platform \cite{geng2020awb}. Such a graph inference costs 2$\times$ FLOPs and 5000$\times$ latency of a 50-layer DNN, ResNet-50, the inference of which on ImageNet requires only 8G FLOPs \cite{canziani2016analysis} and a latency of less than 50 milliseconds \cite{res50_latency}.

To alleviate the aforementioned challenges for unleashing many of GCNs' exciting applications, pioneering works have explored from either the algorithm or hardware level. 
From the algorithm level, commonly used compression techniques have been applied to GCNs,
such as GCN quantization \cite{tailor2020degree} and sparsification \cite{li2020sgcn}.
From the hardware level, most GCN accelerators aim to design innovative micro-architectures and dataflows to boost the acceleration efficiency of the irregular aggregation phase driven by the fact that the acceleration bottleneck of the associated highly sparse and irregular adjacency matrices. 
For example, AWB-GCN \cite{geng2020awb} leverages three auto-tuning techniques to dynamically balance the workload for all processing elements (PEs) to boost the efficiency. 
HyGCN \cite{yan2020hygcn} proposes a window sliding method to improve the locality of non-zero elements in GCNs' adjacency matrices, and leverages intra-vertex/node parallelism for aggregation and weight reuses for combination, respectively.

Despite their impressive performance, GCNs' acceleration efficiency is still limiting, impeding the unfolding of GCNs' great potential in many real-world applications. In this work, we advocate GCN algorithm and accelerator co-design, and make the following contributions:  

\begin{itemize}
    
    \item We propose a \textbf{G}CN algorithm and accelerator \textbf{Co}-\textbf{D}esign framework dubbed GCoD which alleviate the aforementioned irregularity of GCN inference at different granularities 
    and largely resolve the bottleneck inefficiency of GCN computing by harmonizing both algorithm and accelerator innovations. To the best of our knowledge, GCoD is the first co-design framework dedicated for efficient GCN acceleration, opening up an exciting perspective for exploring much more efficient GCN solutions.

    \item On the algorithm level, GCoD integrates a split and conquer training strategy to polarize the graphs to be either denser or sparser in local neighborhoods without compromising the model accuracy, resulting in adjacency matrices that have two levels of workload and enjoys largely enhanced regularity and thus ease of acceleration. In this way, GCNs still preserve large degrees of irregularity (and thus high inference accuracy) while enabling regular data accesses and processes within each workload, favoring hardware efficiency and overall utilization.

\item On the hardware level, GCoD integrates a dedicated two-pronged accelerator to leverage GCoD algorithm's resulting graph adjacency matrices for further boosting the acceleration efficiency. 
\hr{Specifically, one branch incorporates a chunk-based micro-architecture to accelerate the polarized denser subgraphs with regular/denser patterns and balanced workloads; while the other branch accelerates (mostly on-chip) the remaining irregular/sparser but largely reduced sparser workloads (a small portion of non-zeros, e.g., 30\% in Cora). Results of the two branches are then aggregated without conflicts. }

  

  \item Benchmarking experiments and ablation studies on five GCN models and six graph datasets consistently validate the effectiveness of our GCoD framework, e.g., GCoD leads to 15286$\times$, 294$\times$, 7.8$\times$, and 2.5$\times$ speedups over CPUs, GPUs, and the existing SOTA GCN accelerators including HyGCN and AWB-GCN, respectively,
  while maintaining the same or an even better accuracy. 

\end{itemize}

\section{Related Works}

\textbf{Graph Convolutional Networks (GCNs).}
GCNs have amazed us for processing non-Euclidean and irregular data structures \cite{zhang2018end}. Recently developed GCNs can be categorized into two groups: spectral and spatial methods. Specifically, spectral methods \cite{kipf2017semi, peng2020learning} model the representation in the graph Fourier transform domain based on eigen-decomposition, which are time-consuming and usually handle the whole graph simultaneously, making it difficult to parallel or scale to large graphs \cite{gao2019graphnas, wu2020comprehensive}. On the other hand, spatial approaches \cite{hamilton2017inductive,simonovsky2017dynamic}, which directly perform the convolution in the graph domain by aggregating the neighbor nodes’ information, have rapidly developed recently. To further improve the performance of spatial GCNs, Veličković et al. \cite{GAT} introduces the attention mechanism to select information which is relatively critical from all inputs; 
Zeng et al. \cite{zeng2019accurate} proposes mini-batch training to improve GCNs' scalability of handling large graphs; and Xu et al. \cite{xu2018how} theoretically formalizes an upper bound for the expressiveness of GCNs. 
GCoD's innovation is general and thus can be applied on top of different GCN algorithms to boost their hardware acceleration efficiency.

\textbf{GCN Compression.} The prohibitive complexity and powerful performance of GCNs have motivated growing interest in GCN compression.
For instance, Tailor et al. \cite{tailor2020degree} for the first time shows the feasibility of adopting 8-bit integer arithmetic representation for GCN inference without sacrificing the classification accuracy; Two concurrent pruning works \cite{li2020sgcn,zheng2020robust} aim to sparsify the graph adjacency matrices; and Ying et al. \cite{ying2018hierarchical} proposes a DiffPool layer to reduce the size of GCN graphs by clustering similar nodes during training and inference. 
Our GCoD's split and conquer training algorithm explores from a new perspective by enforcing GCNs to naturally present the desired patterns that are potentially hardware friendly and efficient.  


\begin{figure*}
    \centering
    \includegraphics[width=0.9\linewidth]{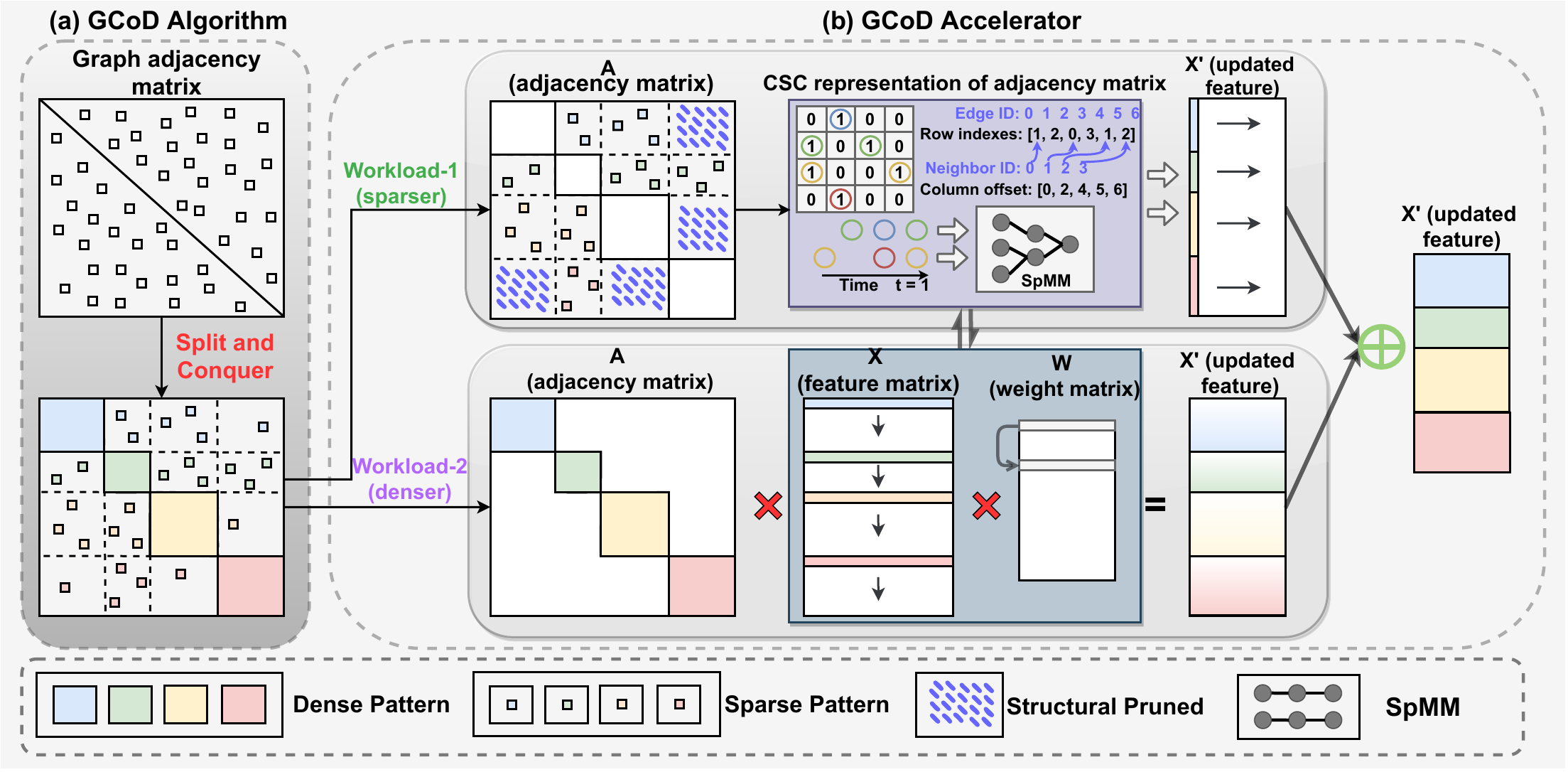}
    \caption{\hr{An overview of the proposed GCoD, an algorithm and accelerator co-design framework dedicated to GCN acceleration.}}
    \vspace{-1.5em}
    \label{fig:overview}
\end{figure*}

\textbf{Graph Reordering.} Graph processing applications are notorious for exhibiting extreme irregularity. Prior techniques on graph reordering \cite{karypis1998fast,arai2016rabbit,azad2017reverse} exploit structural properties of real-world graphs to enhance locality, \textit{after training} the graph.
Different from existing reordering works, our GCoD algorithm is dedicated for hardware-friendly GCN inference and integrates both pruning and reordering \textit{during GCN training} to enforce polarization and hardware-aware sparsification. As such, the resulting GCNs enjoy a higher degree of data locality while maintaining or even improving the accuracy. Specifically, our GCoD algorithm leads to a SOTA graph pruning ratio (10\% edge removal \cite{li2020sgcn}) without accuracy degradation and two distinct workloads that are explicitly leveraged by our GCoD accelerator.


\textbf{GCN Inference Acceleration.}
The ultra-sparsity in graph adjacency matrices (e.g., Pubmed dataset has 99.989\% sparsity vs. sparse DNNs have 88.9 $\sim$ 92.3\% sparsity \cite{han2015learning}) requires dynamic and irregular data accesses for GCNs' feature aggregation, which has a different execution pattern from DNN accelerators and thus motivates dedicated GCNs accelerators \cite{autenhardware,liang2020engn}.
HyGCN \cite{yan2020hygcn} characterizes the hybrid execution patterns for exploring both the intra/inter-vertex parallelisms to handle the irregularity in the aggregation phase and reusability in the combination phase, respectively.
Later, AWB-GCN \cite{geng2020awb} identifies the workload imbalance problem in the aggregation phase, the non-zero values (i.e., connected neighbors) in adjacency matrices are regionally clustered, and proposes autotuning workload balancing techniques to alleviate the runtime imbalance.
Another trend is to summarize the design space of the dataflow and micro-architecture optimization in GCN accelerators \cite{garg2021taxonomy}, and provide automated framework to generate suitable hardware for the given GCN applications \cite{liang2020deepburning,DBLP:journals/corr/abs-2109-08983}. For example, G-CoS \cite{DBLP:journals/corr/abs-2109-08983} develops the first co-search framework that can automatically search for the matched GNN structures and accelerators to maximize both task accuracy and acceleration efficiency. 
In contrast,
Our GCoD explores dedicated algorithm-accelerator co-design for GCN inference in order to further boost the overall utilization and acceleration efficiency thanks to the enforced regularity.
\section{GCoD: Motivation \& Overview}
\label{sec:GCoD}

\textbf{Why GCN Inference Is Inefficient.}
There exists a fundamental dilemma associated with GCN inference acceleration: To accelerate GCN inference, the irregularity of GCNs' adjacency matrices need to be reduced, which can inevitably degrade the inference accuracy; On the other hand, maintaining GCNs' irregularity and thus their excellent accuracy can lead to extremely high hardware costs of GCN inference as demonstrated in recent works \cite{yan2020hygcn,geng2020awb,liang2020engn,liang2020deepburning,kiningham2020grip,chen2020rubik,garg2021taxonomy,zeng2020graphact}; both limiting their more extensive applications. 

\textbf{Why GCoD Boosts GCN Inference Efficiency.}
Fig. \ref{fig:overview} shows an overview of the proposed GCoD framework, which resolves the above dilemma in a clever way.
We first leverage a split and conquer training algorithm to polarize the graph to be either denser or sparser, and then design a dedicated two-pronged accelerator for separately handling each of the resulting two levels of workload.
The motivating intuition is very simple: If the mass irregularity is clustered into different classes with each requiring similar data access and process patterns and being handled by a dedicated sub-accelerator, then the above dilemma can be largely resolved when dedicating one sub-accelerator to process one of the aforementioned two workloads. In this way, GCNs still preserve their advantageous large degrees of irregularity (and thus inference accuracy) while enabling mostly regular data accesses and processes within each workload, favoring potential hardware efficiency and overall utilization. Next we will introduce our GCoD algorithm and accelerator.


\section{The Proposed GCoD Algorithm}

In this section, we present the proposed GCoD algorithm. Specifically, we first present the preliminaries of GCN training and graph optimization in Sec. \ref{sec:alg_prepare}, and then introduce the design considerations, detailed GCoD algorithm formulation, and the early-stop efficient training pipeline in Sec. \ref{sec:algorithm}.

\subsection{Preliminaries of GCNs and Graph Optimization}
\label{sec:alg_prepare}
\textbf{GCN Notation and Formulation.} Let $\mathcal{G} = (V, E)$ represents a GCN graph, where $v_i \in V$ and $(v_i, v_j) \in E$  denote the nodes and edges, respectively; and $N = | V |$ and $M = | E |$ denote the total number of nodes and edges, respectively. The node degrees are denoted as $d = \{d_1, d_2, \cdots, d_N\}$ where $d_i$ indicates the number of neighbors connected to the node $v_i$. We define $D$ as the degree matrix whose diagonal elements are formed using $d$. Given the adjacency matrix $A$ and the feature matrix $X = \{x_1, x_2, \cdots, x_N\}$ of the graph $G$, a two-layer GCN model \cite{kipf2017semi} can then be formulated as:
\begin{equation}\label{eq:gcn}
    Z = f(A, X) = \text{softmax} \left(\hat{A} \,\, \text{ReLU} \left(\hat{A}XW_0\right) W_1 \right),
\end{equation}
where $\hat{A}$ is a normalized version of $A$: $\hat{A}=D^{-\frac{1}{2}} A D^{-\frac{1}{2}}$. 
The whole GCN inference can be viewed as two separated phases: \textit{Aggregation} and \textit{Combination}.

\begin{itemize}[leftmargin=5mm,topsep=0mm]
\itemsep -0.3\parsep
    \item \textit{Aggregation}: For each node in the graph, a GCN aggregates its 1-hop neighbor nodes' feature vectors into a unified feature vector, which corresponds to the multiplication of the adjacency matrix and feature matrix $\hat{A} X$.
    \item \textit{Combination}: The aggregated feature vector will be further transformed to another feature vector using an MLP network (shared between nodes) for learning better representations, which corresponds to the multiplication between the feature matrix and weight matrix, i.e., $X W$.
\end{itemize}
After the feature vectors update, a softmax function is applied in a row-wise manner, i.e., $\textit{softmax}(x_i) = \text{exp} (x_i) / \sum_i \text{exp} (x_i)$  \cite{kipf2017semi}. For semi-supervised multiclass classification, the loss function captures the cross-entropy errors over all labeled examples:
\begin{equation}\label{eq:gcn_loss}
    \mathcal{L}_{GCN}(W) = - \sum_{n \in \mathcal{Y}_N} \sum_f Y_{nf} \, ln(Z_{nf}),
\end{equation}
where $\mathcal{Y}_N$ is the set of node indices that have labels, $Y_{nf}$ is the ground truth label matrix, and $Z_{nf}$ denotes the predicted possibilities of node $n$ belonging to class $f$. During GCN training, $W_0$ and $W_1$ are updated via gradient descents.

    
    
    

\textbf{Graph Optimization.} The goal of graph optimization is to enforce the graph adjacency matrix for achieving desired patterns, which usually improve the graph's regularity so that the GCN models can be more hardware friendly on their target platforms.
For example, graph sparsification aims to reduce the total number of edges in graphs (i.e., the size of the adjacency matrix).
A SOTA graph optimization pipeline \cite{li2020sgcn} is to first pretrain GCNs on their full graphs, and then optimize the graphs based on the pretrained GCNs. Note that the weights of GCNs are not updated during graph optimization, during which $W$ is replaced with $A$ in Eq. (\ref{eq:gcn_loss}) to derive the loss function $\mathcal{L}_{GCN}(A)$. As such, the overall loss function during graph optimization can be written as \cite{li2020sgcn}:
\begin{equation} \label{eq:graph_loss}
    \mathcal{L}_{Graph}(A) = \mathcal{L}_{GCN}(A) + \mathcal{L}_{Reg}(A),
\end{equation}
where $\mathcal{L}_{Reg}$ denotes the regularization term. Such a graph optimization enables practitioners to enforce regularized patterns in graphs for achieving more efficient GCN inference.

\subsection{The Proposed GCoD Algorithm}
\label{sec:algorithm}

\subsubsection{GCoD: Split and Conquer Algorithm}
\label{sec:GCoD_algorithm}

\textbf{Split and Conquer Chunk Design.}
Our GCoD algorithm alleviates the ultra-high sparsity (e.g., 99.989\% in the Pubmed dataset) and irregularity in GCNs' adjacency matrices by leveraging \textit{subgraph classification} to enforce regularity at both coarse- and fine-grained granularities, and adopt \textit{group partitioning} to further boost the processing efficiency:

\begin{figure}
    \centering
    \includegraphics[width=\linewidth]{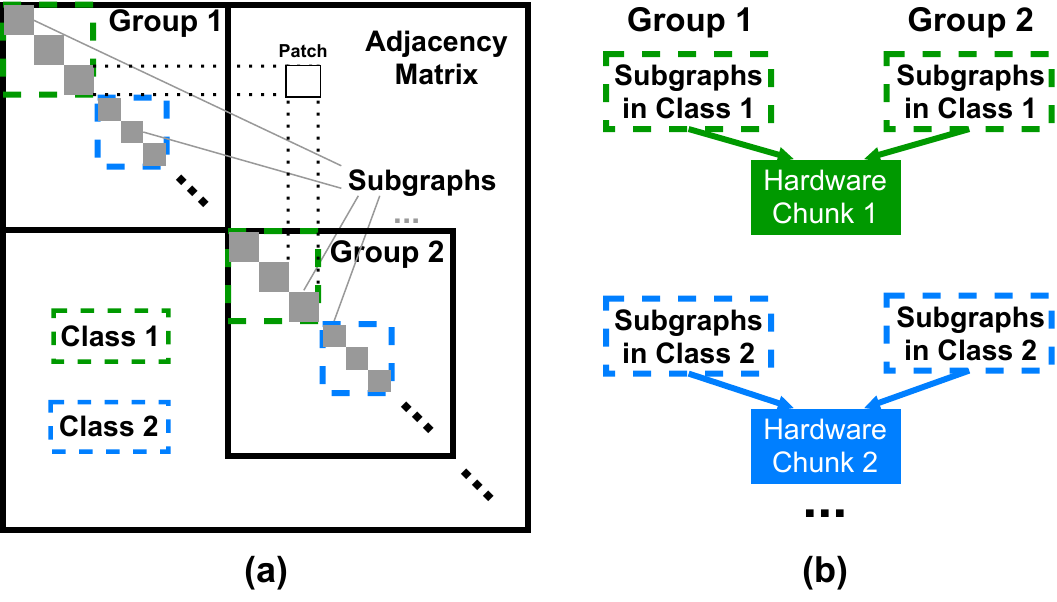}
       \caption{Illustrating (a) GCoD's defined \textit{group}, \textit{class}, and \textit{subgraph} within GCNs' graph adjacency matrices, where nodes with similar degrees are categorized into the same class, 
       each class is further divided into subgraphs with a similar number of edges,
       and all the subgraphs within the same class are evenly distributed into different groups, and (b) each hardware chunk (i.e., sub-accelerator) handles the same kind of classes from all the groups.}
    \label{fig:chunk_design}
    \vspace{-.5em}
\end{figure}

\begin{itemize}[leftmargin=5mm,topsep=0mm]
\itemsep -0.3\parsep
    \item \textit{Subgraph Classification}: To reduce the irregularity of GCNs' graph adjacency matrices,
    we cluster nodes with similar degrees into the same class (different classes are denoted using dashed boxes in Fig. \ref{fig:chunk_design}).
    To further achieve a finer-grained regularity within each class,
    we further divide each class into subgraphs with each having a similar number of edges. 
    Therefore, subgraphs within the same class share balanced workloads and favor regular and thus efficient hardware acceleration, which are processed using one sub-accelerator (i.e., chunk) in GCoD.
    Sub-accelerators have unique hardware resources dedicated to handle the workload patterns of the corresponding subgraphs, and can process in parallel.
    \item \textit{Group Partitioning}: 
    We uniformly distribute subgraphs within the same class into different groups.
   \hr{Such group partitioning reduces the boundary connections to enforce the sparser patterns. Therefore, we can treat all of the sparser patterns as one unique workload (w.r.t. a sub-accelerator), which simplifies hardware designs and the communication among different sub-accelerators.}
\end{itemize}

The resulting patterns' irregularity in GCNs' graph adjacency matrices still maintain relatively high, while each sub-accelerator can process the same class of nodes 
having similar degrees and thus similar data access and process workloads, which can also be well supported by our dedicated accelerator (see Sec. \ref{sec:GCoD_accclerator}) as well as SOTA chunk-based accelerator architectures \cite{shen2017maximizing}. Next, we introduce GCoD training pipeline to enforce the above patterns, and its further structural removal of graph connections to increase structured sparsity.

\begin{wrapfigure}{r}{0.25\textwidth}
    \vspace{-0.12in}
    \centering
    \includegraphics[width=0.23\textwidth]{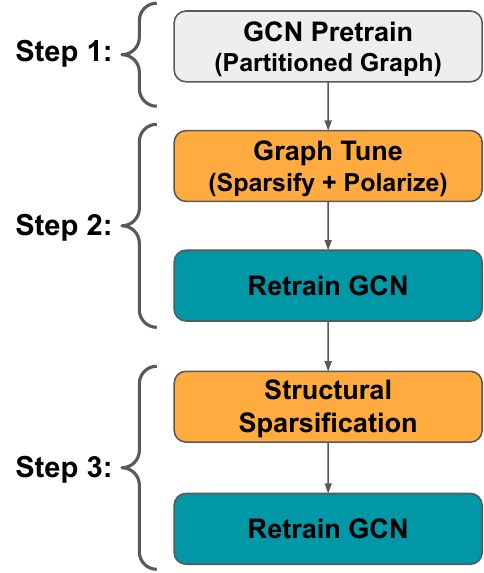}
    \caption{\hr{GCoD training flow.}}
    \vspace{-1em}
    \label{fig:HA-GCN_train}
\end{wrapfigure}

\textbf{Training Pipeline.}
As shown in Fig. \ref{fig:HA-GCN_train}, the training pipeline of GCoD can be divided into three steps: (1) pretraining GCNs on the partitioned graphs; (2) tuning 
\hr{(i.e., sparsify and polarize)}
the graphs based on the pretrained GCN model; and (3) structural sparsify graph adjacency matrices. 
In Steps (2)-(3), GCN retraining is needed to restore the test accuracy after tuning or sparsification, where Step (2) 
can be iterated several times to maximize the sparsity while maintaining the accuracy.
Next, we furhter elaborate each step, including the \textit{graph partitioning}, 
\hr{\textit{sparsify and polarize},} 
and \textit{structural sparsification}, of Steps (1)-(3).

\begin{figure}[t]
    \centering
    \includegraphics[width=\linewidth]{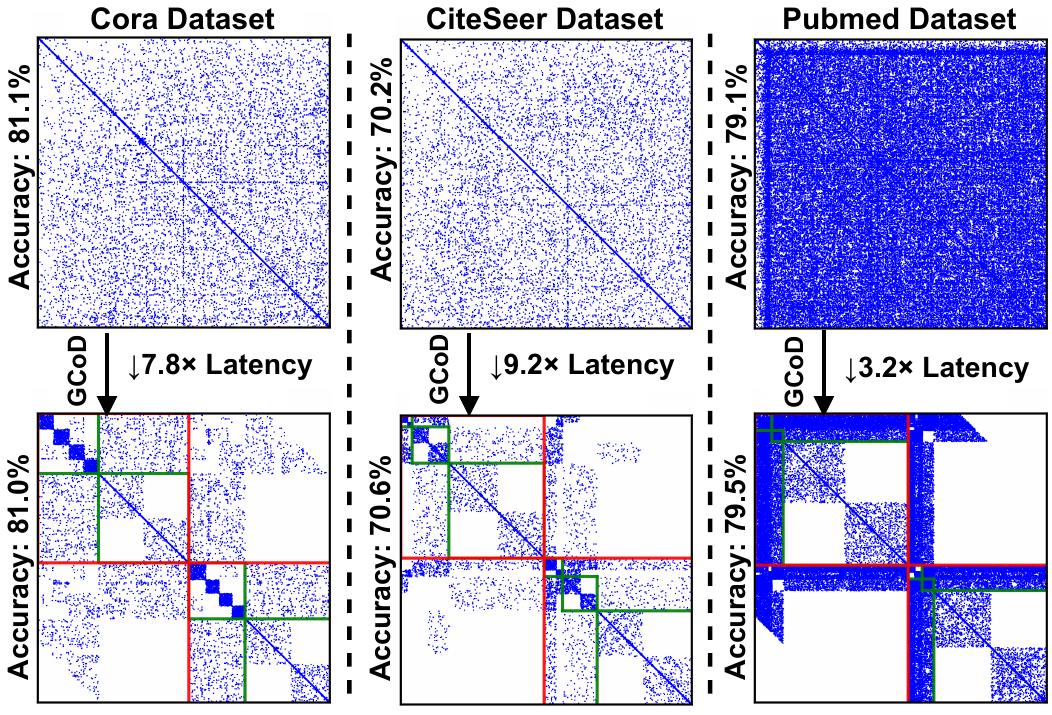}
       \caption{Visualizing three datasets' graph adjacency matrices \textbf{before and after} applying our GCoD algorithm, where green lines separate subgraph classes while red lines partition the groups. Note that non-zero dots in the matrices are enlarged for better visualization effects.}
    \label{fig:vis_chunk}
    \vspace{-1.5em}
\end{figure}

\begin{itemize}[leftmargin=5mm,topsep=0mm]
\itemsep -0.3\parsep
    \item \textit{Step 1: Graph Parititioning.}
    Suppose we aim to separate a graph into $G$ groups and each group contains $C$ subgraph classes. To partition the graph into $S$ subgraphs, we first extract $C$ subgraph classes $\mathcal{G}[c] = \{i \mid \hat{d}_{c-1}\leq d_i < \hat{d}_{c}\}$ for $ 1\leq c\leq C $, with $d_i$ representing the in-degree of the $i$-th node which will be divided to the $c$-th class, if $\hat{d}_{c-1}\leq d_i < \hat{d}_{c}$. Note that $\hat{d}_c$ belongs to the degree partition list which is predefined as $0 = \hat{d}_0 < \cdots < \hat{d}_{C}=\infty$. In this way, all nodes in the same class $\mathcal{G}[c]$ share similar degrees.
    Next, we use METIS \cite{karypis1998fast} to partition each class $\mathcal{G}[c]$ into workload balanced subgraphs. Finally, the $S$ subgraphs are then uniformly distributed across $G$ groups.
    
    \item \hr{\textit{Step 2: Sparsify and Polarize.}
    The goal of graph sparsification and 
    polarization
    is to reduce the total number of non-zero values within the adjacency matrices (i.e., the edges in GCNs' graphs) and 
    polarize the adjacency matrices towards denser and sparser patterns, respectively.
    Note that the weights of GCNs are not updated during this step. Therefore, $W$ is replaced with $A$ in Eq. (\ref{eq:gcn_loss}) to derive the loss function $\mathcal{L}_{GCN}(A)$.
    In summary, the overall loss function for this step can be written as:
    \vspace{-0.5em}
    \begin{equation}\label{eq:tuning}
         \mathcal{L}_{Graph}(A) = \mathcal{L}_{GCN}(A) + \mathcal{L}_{SP}(A) + \mathcal{L}_{Pola}(A),
    \end{equation}
    where $\mathcal{L}_{SP}$ and $\mathcal{L}_{Pola}$ denote the sparse regularization and 
    polarization terms, respectively. Ideally, $\mathcal{L}_{SP}$ will become zero if the sparsity of the graph adjacency matrices reaches the target pruning ratio (e.g., $\nicefrac{\|A_{prune}\|_0}{\|A\|_0} \leq 1 - p$ for a given ratio of $p$);
    $\mathcal{L}_{Pola}$ equals to $\nicefrac{1}{M} \cdot \sum \|i - j\|$, where $M$ denotes the total number of non-zeros in the adjacency matrices, and $i, j$ represent the $x$- and $y$-coordinate values of all non-zero elements within the adjacency matrices $A$.
    As the overall loss function $\mathcal{L}_{Graph}(A)$ is not differential, we follow \cite{li2020sgcn} to use ADMM for minimizing the non-differential loss function using gradient descent.}

    \item \textit{Step 3: Structural Sparsification.} To further improve GCNs' inference efficiency, GCoD algorithm further leverages the patch-based structural sparsity within the graph adjacency matrices, where the patch definition is illustrated in Fig. \ref{fig:chunk_design}. Specifically, GCoD prunes patches of which the number of non-zero elements is smaller than a specified threshold $\eta$, which balances the resulting sparsity and achieved GCN accuracy. In this work, $\eta$ ranges from 10 to 30 for different graph datasets.
    As a results, the final optimized graph adjacency matrices will have some vacancies as shown in Fig. \ref{fig:vis_chunk}.
\end{itemize}

\textbf{Visualization.} We visualize the graph adjacency matrices before and after GCoD training in Fig. \ref{fig:vis_chunk}, which clearly demonstrates the improved regularity and the effectiveness of GCoD split and conquer training strategy: achieving on average 7.8$\times$ speedups  over HyGCN \cite{yan2020hygcn} when evaluating on the proposed accelerator architecture in Sec. \ref{sec:GCoD_accclerator}, while maintaining or even improving the test accuracy.

\subsubsection{Efficient Training Pipeline via Early-stopping}
\label{sec:eb_training}
The vanilla training algorithm of GCoD enables the opportunity of boosting GCN inference efficiency yet requires a nontrivial training overhead as compared to the standard GCN training.
To reduce its training overhead, we propose to first identify the winning subnetworks from the origin GCN networks at the very early training stages (e.g., 10$\sim$20 epochs over a total of 400 training epochs) following \cite{You2020Drawing,you2021gebt}, resulting in much improved training efficiency in both Step 1 (via early-stopping) and Steps 2-3 (via only retraining GCN subnetworks) without compromising the final accuracy. 
In this way, our GCoD algorithm only requires a comparable or even a lower training cost \textbf{(0.7$\times$ $\sim$ 1.1$\times$)} than the standard GCN training. In particular, GCoD training algorithm leads to at most 10\% training overhead when evaluated on SOTA five models and six datasets.
The three steps in GCoD training algorithm account for about 5\%/50\%/45\% of the total training cost, respectively, where the dominated steps 2-3 require to retrain the GCN subnetworks from scratch.


\section{The Proposed GCoD Accelerator}
\label{sec:GCoD_accclerator}

\subsection{Motivation for GCoD Accelerator}
\label{sec:motivation}
\textbf{Opportunity.}
Our proposed GCoD algorithm exhibits a great potential in alleviating the irregularity in GCNs' graph adjacency matrices. However, this potential cannot be fully exploited by existing GCN accelerators \cite{yan2020hygcn,geng2020awb} due to 
(1) the resulting two distinct workloads from GCoD algorithm, i.e., the sparser and denser branches as shown in Fig. \ref{fig:overview}; and
(2) the lack of opportunities in existing GCN accelerators to dedicate for different classes (w.r.t. similar node degrees and balanced workloads within each of them) in the denser branch, and to fully leverage the reduced workloads and enforced structural sparsity in the sparser branch. As such, GCoD accelerator is motivated to take advantages of the new opportunities resulting from GCoD algorithm to further boost the acceleration efficiency.

\begin{figure}[t]
    \centering
    \includegraphics[width=\linewidth]{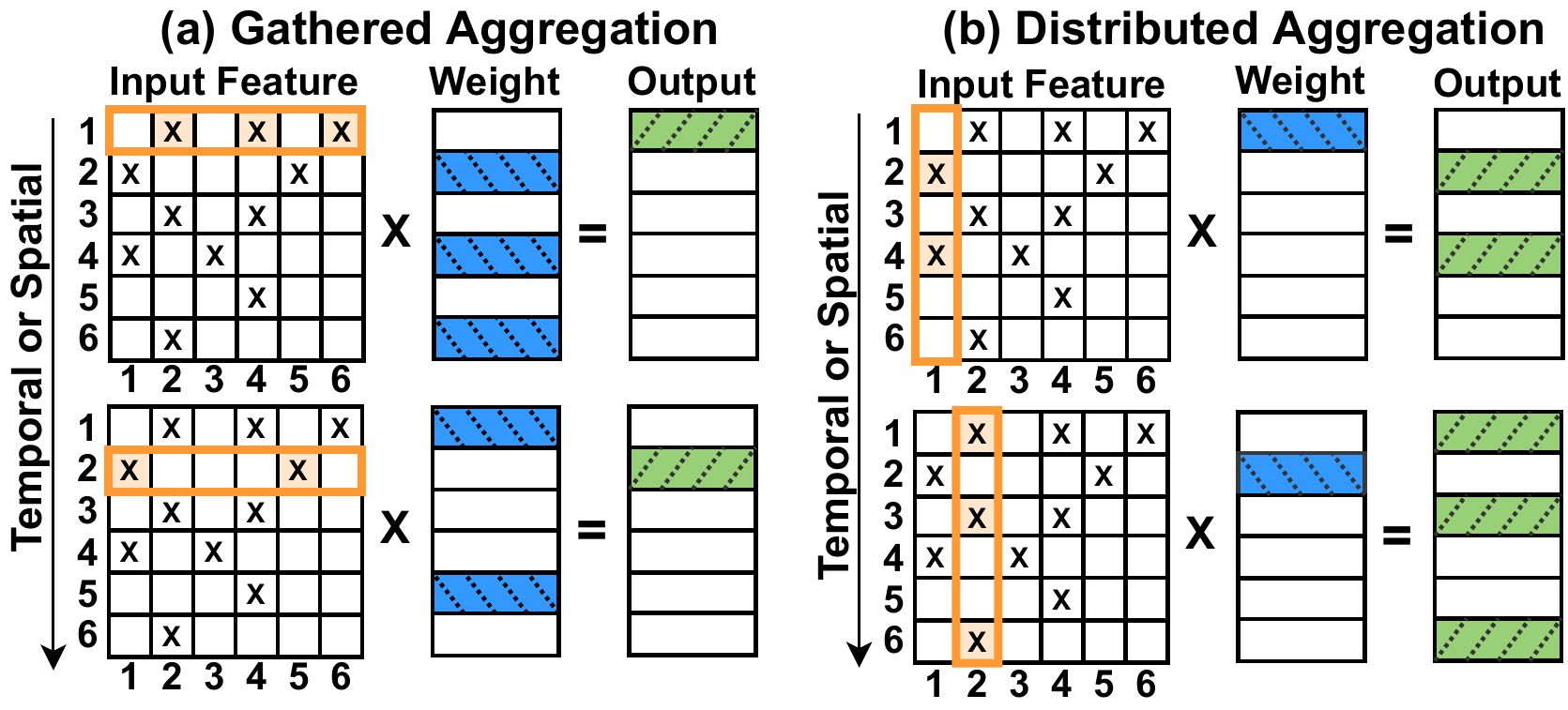}
    \caption{\hr{Illustrating the \textit{gathered} and \textit{distributed} aggregations.}}
    \vspace{-1em}
    \label{fig:motivation}
\end{figure}

\textbf{Design Exploration.} Here we first discuss the two typical designs in existing GCN accelerators for accelerating the aggregation phase, i.e., the performance bottleneck of GCN processing, 
and elaborate their advantages and disadvantages.
To handle the dominant aggregation phase, SOTA GCN accelerators either adopt \textit{gathered} aggregation or \textit{distributed} aggregation. In particular, \textbf{gathered aggregation} (see Fig. \ref{fig:motivation} (a)) executes nodes in a sequential manner, where the neighbor features of each node are gathered in parallel for aggregation. The advantage is that it requires only a small buffer for handling the aggregation results thanks to the good reuses of the intermediate aggregation outputs. However, such gathered aggregation causes irregular and frequent (off-chip) accesses of the 
weights and features, which is often too large to be stored on-chip, due to the sparse and random distribution of non-zeros in the adjacent matrix. For example, HyGCN \cite{yan2020hygcn} adopts such gathered aggregation.
On the other hand, the latter, i.e., \textbf{distributed aggregation} (see Fig. \ref{fig:motivation} (b)), executes nodes distributively in a parallel manner, where the neighbor features of each node are gathered sequentially for aggregation. Its advantage is that the weight features can be fully reused, because it processes the non-zero elements of the adjacent matrix in a column-wise manner and thus allows the rows of the weight matrix to be reused by all the elements of the same column in the adjacent matrix. This advantage yet comes at a cost of requiring a large buffer to hold the intermediate aggregation results, which is often too large to be stored on-chip and thus leads to frequent off-chip accesses.
Furthermore, with the often varying number of neighbors for different nodes, the required off-chip memory accesses can be rather irregular and thus leads to further workload imbalance and inefficiency.
For example, AWB-GCN \cite{geng2020awb} adopts such distributed aggregation. 


In general, the gathered aggregation needs more off-chip bandwidth to access weights and features while the distributed aggregation needs more on-chip storage to hold the aggregation results. 
Given that bandwidth is more limited when processing GCNs which have a poor data locality and extreme irregularity, the distributed aggregation better favors GCN acceleration efficiency as compared to the gathered aggregation, as verified by the much improved performance offered by AWB-GCN (distributed) over HyGCN (gathered). As such, GCoD considers the distributed aggregation design, and strives to further alleviate the associated workload imbalance problem and to reduce on-chip storage requirements for much boosted GCN inference efficiency. 
In particular, GCoD algorithm training enforces two distinct workloads for the aggregation phase without compromising the accuracy. 
Each of the resulting two workloads is now better suited for overcoming distributed aggregation's disadvantages.
We next present the detailed micro-architecture design in Sec. \ref{sec:architecture}.

\vspace{-0.5em}
\subsection{GCoD Accelerator's Micro-architecture}
\label{sec:architecture}

\begin{figure}[t]
    \centering
    \includegraphics[width=\linewidth]{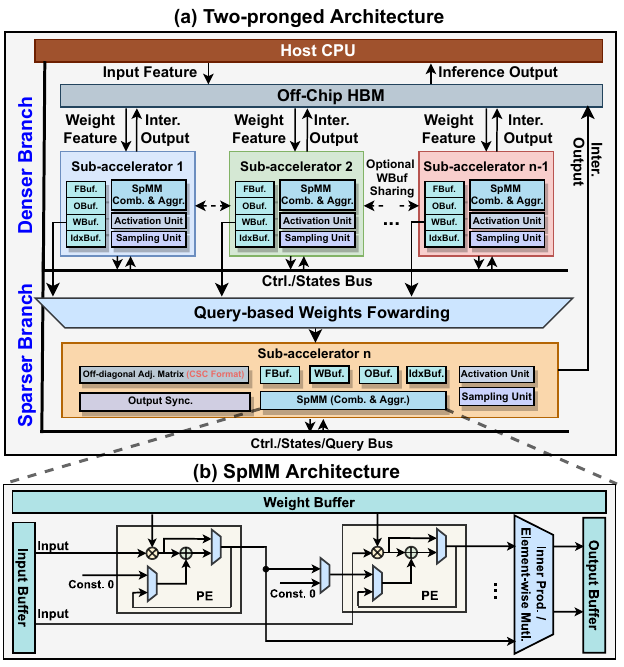}
    \caption{\hr{Illustrating the micro-architecture of GCoD accelerator.}}
    \vspace{-1.2em}
    \label{fig:Architecture}
\end{figure}

\textbf{Architecture Overview.}
Fig. \ref{fig:Architecture} illustrates the overall micro-architecture of the GCoD accelerator.  For better processing elements (PEs) utilization and reduced off-chip memory access during the performance dominant aggregation phase, 
\hr{GCoD accelerator consists of two separate computing branches with each dedicated to process 
the denser workload and sparser workload
of GCoD algorithm's resulting adjacency matrices, respectively.} 
As shown in the upper part of Fig. \ref{fig:Architecture}, 
\textbf{the Denser Branch} 
processes the enforced regular dense subgraphs along the diagonal line of the adjacency matrices with an array of parallel sub-accelerators to handle the more intense workload while maintaining the workload balancing through complexity proportional resource allocation among the sub-accelerators.
\textbf{The Sparser Branch}, 
as shown in the bottom part of Fig. \ref{fig:Architecture},
at the same time handles the remaining irregular but significantly lightweight 
\hr{sparser}
workloads mostly \textit{on-chip}, through the Compressed Sparse Column (CSC) input format and the proposed weight forwarding technique, largely avoiding frequent and large volume data movements from the off-chip memory.
Each sub-accelerator within the branches is capable of processing the task of sparse-dense matrix multiplication (SpMM), non-linear activation, and node sampling. We term the proposed micro-architecture comprising two separate computing branches as \textit{two-pronged architecture} hereafter. 
The characteristics of each branch are summarized in Tab. \ref{tab:summary}. 
For minimal controlling overhead and interruption to the computing flow, these two branches are equipped with separate output buffers, so that their generated results can be written into the buffers in parallel and are further synchronized and accumulated to produce the final outputs, as illustrated in Fig~\ref{fig:overview}.
We next elaborate the design of each branch.

\begin{table}[t]
  \centering
  \caption{Summarizing the characteristics of GCoD accelerator's Denser and Sparser Branches.}
  \resizebox{0.49\textwidth}{!}{
    \begin{tabular}{lccccccc}
    \hline
    \multicolumn{1}{c}{} &\textbf{\tabincell{c}{Multi\\Chunks}} & \textbf{\tabincell{c}{On-chip\\Storage}} & \textbf{\tabincell{c}{Off-chip\\Access}} & \textbf{\tabincell{c}{Arch.\\Reuse}} & \textbf{\tabincell{c}{Data\\Reuse}} & \textbf{Workloads} \\
    \hline
    \hline
    \tabincell{l}{W/o GCoD} & N & H & H & N & N  & \textbf{\tabincell{c}{Heavy \& \\ Imbalanced}} \\
    \hline
    GCoD Denser & Y & L & L & Y & Y  & \textbf{Balanced}\\
    GCoD Sparser & N & H & L & Y & Y  & \textbf{Light} \\
    \hline
    \end{tabular}%
  }
  \label{tab:summary}%
  \vspace{-1em}
\end{table}%

\textbf{Denser Branch.}
As shown in Sec. \ref{sec:algorithm}, subgraphs from different classes will have different dimensions, resulting in different workload sizes when being processed. Thus, to balance the workload of processing the subgraphs from different classes and meanwhile maintain a high degree of parallelism, we adopt a chunk-based architecture for the denser branch acceleration,
where sub-accelerators process the subgraphs from different classes simultaneously. To achieve balanced workload, we first estimate the workload size for the subgraphs within each class and assign each sub-accelerator to a different class. After that, given the available hardware resource budget, we allocate hardware resource to each sub-accelerator proportional to its assigned subgraphs' workload size. In particular, (1) for PEs allocation, we use the number of multiply and accumulate operations (MACs) with sparsity considered to characterize each workload's size, and assign each sub-accelerator the number of PEs proportional to the assigned subgraphs' number of MACs; (2) For on-chip memory and off-chip bandwidth allocation, we first calculate all the input/output feature maps and weights sizes when processing the subgraphs within each class, and then similarly assign memory and bandwidth to each sub-accelerator proportional to the sum of the calculated feature maps and weights sizes. 
As the subgraphs within each class have similar workload sizes, benefiting from GCoD algorithm's integrated subgraph classification, GCoD accelerator adopts the same sub-accelerator assigned to each class to process the subgraphs within the same class.
For convenience, hereafter we term a sub-accelerator for processing workload balanced subgraphs within the same class as a ``chunk'',  as illustrated in Fig. \ref{fig:Architecture} (a). Note that the number of chunks equals to the number of classes and is obtained during GCoD algorithm's training.  Throughout the densor branch, either dense or Coordinate (COO) format inputs and weights are assumed for reduced controlling overhead.

It is worth noting that such a chunk-based accelerator design naturally achieves workload balance without the necessity of on-the-fly autotuning, differentiating itself from previous SOTA accelerators, e.g., AWB-GCN \cite{geng2020awb}.

\textbf{Sparser Branch.} Benefiting from GCoD algorithm, the off-diagonal workloads feature much reduced data and computation density (50\% $\sim$ 75\%), 
as shown in Fig. \ref{fig:vis_chunk}. As such, GCoD accelerator handles the data movements associated with these workloads, i.e., when processing the workload in the sparser branch, with (mostly) the on-chip memory by utilizing (1) a CSC data format for the input data (i.e., entries in the adjacency matrices) and (2) query-based weight (i.e., features to be aggregated) forwarding. 
\begin{itemize}
    \setlength{\itemsep}{0pt}
    \setlength{\parsep}{0pt}
    \item \textit{CSC Format Inputs}: By storing in the CSC data format, GCoD accelerator is able to fit most of the input data to the on-chip buffer, thanks to the drastically reduced adjacency matrices' density and CSC's smaller storage overhead as compared with the COO format. To be compatible with the CSC format and the subsequent weight forwarding technique, GCoD accelerator adopts the distributed aggregation design (see Fig. \ref{fig:motivation} (b)), which consumes column(s) of the inputs every clock cycle.
    
    \item \textit{Query-based Weight Forwarding}: Because the denser and sparser branches operate in parallel, when the sparser branch is working on certain columns of the input data, it is likely that some chunks in the denser branch is working on the same columns but different rows, thus sharing the same needed rows of weights as the sparser branch. Therefore, for the weights needed during the sparser branch, instead of loading them from the off-chip memory, the sub-accelerator in the sparser branch will query the weight buffers of the corresponding chunks in the denser branch for accessing the already loaded weights. 
    \hr{Specifically, the weight forwarding is performed on the demand of the sparser branch.  The sparser branch sub-accelerator first determines which of the other sub-accelerators to query based on the queried weights’ row indices which are also the adjacency matrix’s column indices. By checking the predefined  location of the queried sub-accelerator’s index buffer, the sparser branch sub-accelerator will acquire the range of the weight data currently stored in the weight buffer. If the queried data falls in the range, its address within the weight buffer will be calculated based on the known range. Then, the sparser branch sub-accelerator will access that specific address for the queried weight data. Overall, for the sparser branch’s weight, about 63\% of the data will be accessed through the query-based weight forwarding. In particular, the sparser branch sub-accelerator works on multiple columns across different classes, simultaneously. The sparser branch sub-accelerator is able to finish the computation at the similar pace to all the parallel sub-accelerators in the denser branches, because of the increased sparsity and predefined resource allocation. The overall matched pace ensures the decent amount of likelihood of weight forwarding. However, because of the further temporal tiling of each class i.e., the buffer may not fit all the weights belonging to the specific class and the denser and sparser branches will not be synchronized until the end of aggregation, weight forwarding conditions will not always be satisfied. For these cases, weights will be instead loaded from the off-chip memory.}
\end{itemize}
\vspace{-0.3em}

\begin{figure*}[t]
    \centering
    \includegraphics[width=\linewidth]{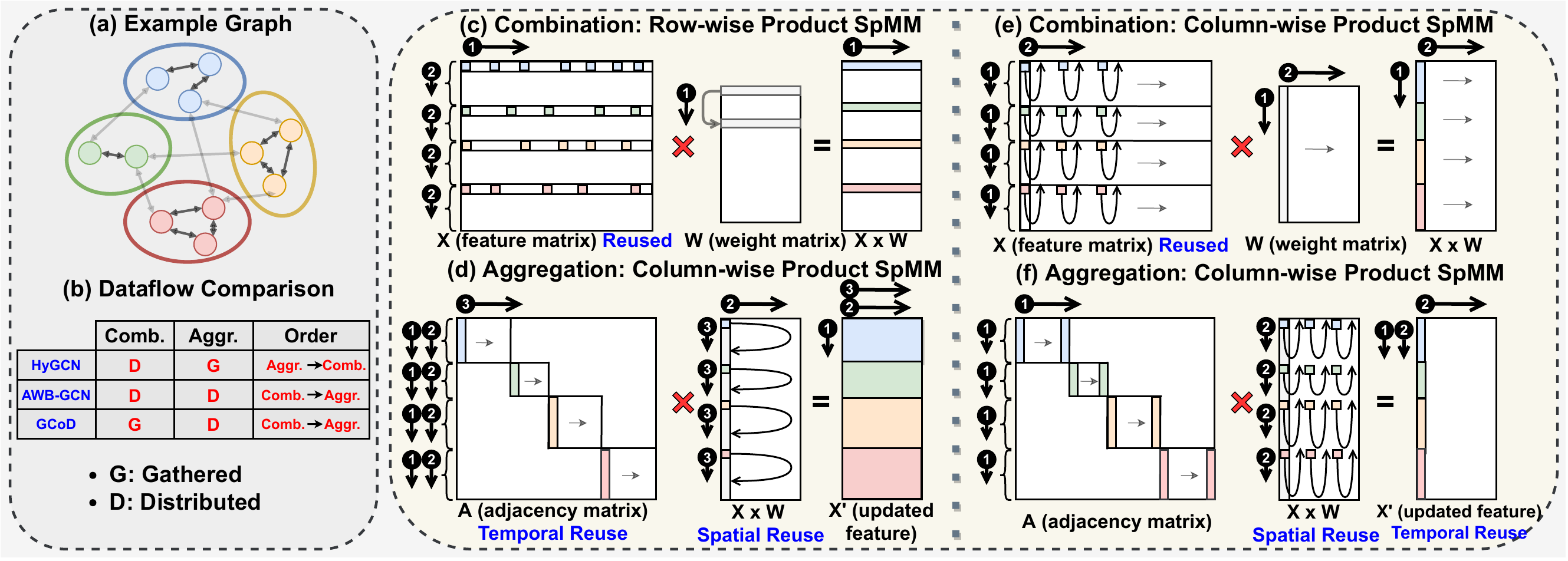}
        \caption{\hr{An illustration of the efficiency- and resource-aware pipelines for various data reuses within our GCoD accelerator, where the execution order is numbered in each sub-figure.}}
    \vspace{-1em}
    \label{fig:order}
\end{figure*}

Our GCoD accelerator adopts a similar architecture design and workload allocation for processing the sparser workloads of the sparser branch as that of the denser branch, but with only one sub-accelerator. For larger graphs scaling to billion-edge levels under which the on-chip storage cannot handle the sparser workloads, our GCoD accelerator will consider fine-grained pipelines as described in Sec. \ref{sec:subaccelerator_arch} to separate the workloads and continue maintaining the less off-chip memory accesses benefits as compared to vanilla graphs.
In addition, the additional structural sparisty (up to 10\%) of GCoD algorithm's adjacency matrices leads to more columns to be entirely skipped and thus facilitates the accumulation of partial results from two branches.

%

\textbf{Sub-accelerator Architecture.}
\label{sec:subaccelerator_arch}
As shown in Fig. \ref{fig:Architecture}, to support different layer dimensions and the special operations from various GCN structures, the sub-accelerators are equipped with multiple functional units:
\vspace{-0.2em}
\begin{itemize}
    \item \textbf{Dedicated Buffers} to favor local reuse opportunities, with sizes decided in the above resource allocation stage. \hr{The buffers are implemented in either block RAM or look-up tables depending on their sizes and required parallel read/write ports.}
    \item \textbf{Sparse/Dense Matrix Multiplication Engine (SpMM)} which supports both dense and sparse matrix multiplication. Specifically, in the dense format, it takes vectors of inputs and weights and performs either inner products or element-wise multiplications between inputs and weights, as shown in Fig.~\ref{fig:Architecture} (b). When sparsity is considered, thanks to the adopted COO input format or the CSC input format (in distributed aggregation mode), only non-zero elements of the inputs and their indices are loaded for calculation. \hr{Specifically, when handling the sparse matrix multiplication in the denser branch or combination phase, we use the COO format for the adjacency/feature matrix storage and the dense format for the weight storage. As such, by nature we can access only the non-zero adjacency/feature matrix’ values and their location indices to support the sparsity. The loaded elements will be piped to the PEs in Fig.~\ref{fig:Architecture} (b) for result calculations. On the other hand, when handling the sparse matrix multiplication in the sparser branch, we use the CSC format for the adjacency matrix for a smaller storage overhead and the dense format for the transformed (combined) features. Column(s) of the non-zero adjacency matrix elements will be loaded at the same time. With the dataflow restricted to the distributed fashion as in Fig.~\ref{fig:motivation} (b), we can produce the rows of outputs only corresponding to the non-zero elements in each adjacency matrix’s column at a time to fully leverage the sparsity.} 
    
    \item \textbf{Element-wise Activation Units} for the non-linear activation operations. \hr{Specifically, we use gating modules for the ReLU and lookup tables to estimate other non-linear activation functions.}
    \item \textbf{Sampling Units} to schedule the node sampling. \hr{Specifically, we implement a linear shift register to randomly pick from non-zero elements from the adjacency matrices' columns.}
\end{itemize}
\vspace{-0.2em}
The overall micro-architecture is reused for GCN aggregation and combination, as both involve mostly matrix multiplications. The sub-accelerator in the sparser branch has an additional output synchronization module to combine its outputs with the denser branch's outputs. To support large GCNs, computation tiling can be achieved by scheduling the input/weight vectors loaded to the sub-accelerator, without modifying the underlying hardware. Each sub-accelerator communicates with an off-chip High Bandwidth Memory (HBM) through direct memory access to increase the access efficiency. 

\textbf{Efficiency- or Resource-aware Pipeline.}
Distributed aggregation helps reduce off-chip memory accesses. GCoD further enables the intermediate results (i.e., one row of $XW$) to be directly utilized by aggregation through performing combination in a row-wise gathered manner.
Such inter-phase pipeline can largely boost the efficiency and thus is termed as \textit{efficiency-aware pipeline}, as illustrated in Fig. \ref{fig:order} (c) and (d).
However, the above efficiency gain is achieved at a cost of requiring a large on-chip accumulation buffer for storing the aggregation results, otherwise the results have to be transferred back to the off-chip memory.
Benefiting from GCoD's split and conquer training, such an on-chip large buffer can be alleviated but is still required for processing billion-edge large graphs.
To tackle this problem,
we proposed a \textit{resource-aware pipeline} where outputs are also temporally reused so that only one column of aggregation outputs needed to be stored on-chip, making it more suitable for handling large graphs. 
We summarize these two pipelines' characteristics in Tab. \ref{tab:summary_pipeline} and compare GCoD's dataflow with prior works in Fig. \ref{fig:order} (b).
Next, we elaborate data reuses and dataflows in each pipeline.
\begin{table}[t]
  \centering
  \caption{Comparison of two inter-phase pipelines, where RW and CW represent row- and column-wise products.}
  \vspace{-0.7em}
  \resizebox{0.49\textwidth}{!}{
    \begin{tabular}{lcccccc}
    \hline
    \textbf{\tabincell{c}{Inter-Phase}} & \textbf{\tabincell{c}{Comb.\\SpMM}} & \textbf{\tabincell{c}{Aggr.\\SpMM}} & \textbf{\tabincell{c}{On-chip\\Storage}} & \textbf{\tabincell{c}{Off-chip\\Access}} & \textbf{\tabincell{c}{Data\\Reuse}} & \textbf{\tabincell{c}{Fit for \\Graphs}} \\
    \hline
    \hline
    Efficiency-aware & RW & CW & H & L & X, XW, A & Medium \\
    Resource-aware & CW & CW & L & L & X, XW, X$'$ & Large \\
    \hline
    \end{tabular}%
  }
  \label{tab:summary_pipeline}%
  \vspace{-1.8em}
\end{table}%


The core idea of the efficiency-aware pipeline is to perform combination in a row-wise gathered manner so that the intermediate results (i.e., one row of $XW$) can be directly utilized by aggregation. 
During combination, GCNs perform SpMM between the features $X$ and the weights $W$.
Since the dense $W$ is much smaller than $X$ (e.g., 300KB vs. 534MB for Reddit), it can be fully stored on-chip and accessed multiple times upon request without compromising the efficiency.
As such, a row-wise product order is adopted (see Fig. \ref{fig:order} (c)).
Within each sub-accelerator, 
each non-zero element in $X$ needs to multiply an entire row of $W$, and once all elements at the same row of $X$ are fully processed, an entire and dense row of $XW$ is calculated.
During aggregation,
for reducing the off-chip memory accesses of the adjacency matrix $A$,
as shown in Fig. \ref{fig:order} (d),
our GCoD accelerator adopts distributed aggregation to largely reuse the resulting matrix $XW$, which is both dense and large (e.g., 114MB for Reddit). 
Within each sub-accelerator, each element of the resulting row of $XW$ multiplies all the non-zeros in the corresponding column of matrix $A$ so that $XW$ can be fully and spatially reused and $A$ can be temporally reused by all elements in the row of $XW$.

For resource-aware pipeline, 
%
During combination,
as illustrated in Fig. \ref{fig:order} (e), the data reuse pattern remains the same as the efficiency-aware pipeline (i.e., reuse $A$) while the execution order changes from row-wise to column-wise so that the intermediate results will be one column of $XW$.
During aggregation,
as illustrated in Fig. \ref{fig:order} (f), within each sub-accelerator, each element of the resulting column of $XW$ multiplies all non-zeros of $A$ so that not only $XW$ but also the outputs can be fully reused. In this way, we only need to store one column of aggregation outputs to the output buffer.

\begin{figure}[t]
    \centering
    \includegraphics[width=\linewidth]{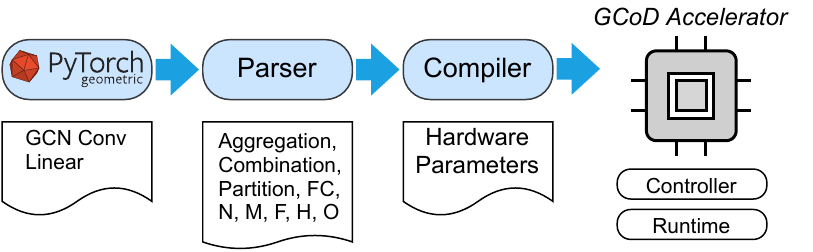}
    \caption{The software-hardware interface pipeline for GCoD.}
    \vspace{-1.8em}
    \label{fig:interface}
\end{figure}

\textbf{Reconfigurability.} 
To support the potential task change after deployment, the proposed GCoD is equiped with a low-cost hardware reconfigurable strategy for use during the hardware compilation process and this compilation cost is a one-time effort for each task. Specifically, a series of C/Verilog based code templates are developed to form the overall hardware architectures with parameterizeable attributes, e.g., number of chunks, PEs, and buffers' sizes. After that, the given GCN will be passed through a network parser to feed the hardware compiler with layer dimensions, e.g., feature size, as illustrated in Fig. \ref{fig:interface}. The hardware compiler will fill in the parameterized attributes in the previous code templates.
The configured hardware architecture will be sent to the platform software, e.g., Vivado~\cite{vivado}, to generate the bitstream for on board deployment. The reconfigurable process cost is amortized across the entire lifetime of each task.

\section{Experiment Results}

In this section, we present a thorough evaluation of the proposed GCoD framework, including the overall benchmark with CPUs/GPUs and SOTA GCN accelerators in Sec. \ref{sec:exp:overall}, and the evaluation and ablation studies of GCoD algorithm and accelerator in Sec. \ref{sec:exp:algorithm} and Sec. \ref{sec:exp:hardware}, respectively.

\begin{figure*}[t]
    \includegraphics[width=\linewidth]{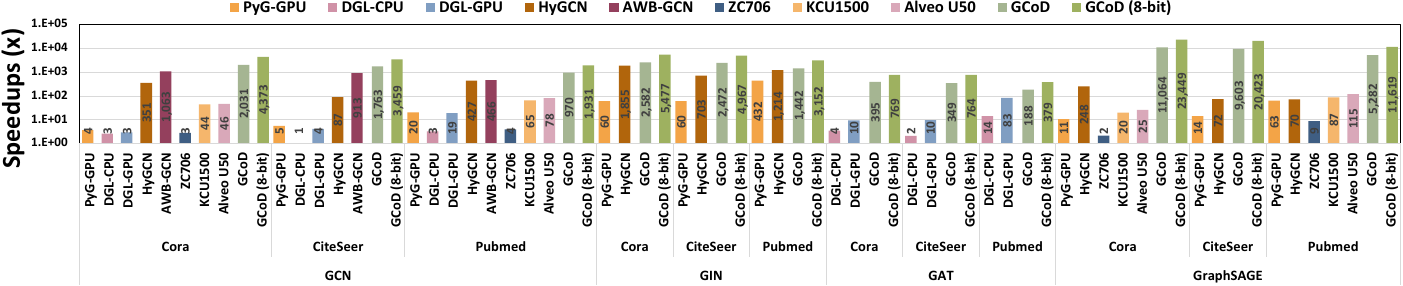}
    \caption{The normalized inference speedups (w.r.t. PyG-CPU) achieved by our GCoD framework over \textbf{nine} SOTA baselines on \textbf{four} variant GCN models and \textbf{three} representative citation graph datasets.}
    \vspace{-0.5em}
    \label{fig:overall_speedup}
\end{figure*}

\begin{figure}[t]
    \includegraphics[width=\linewidth]{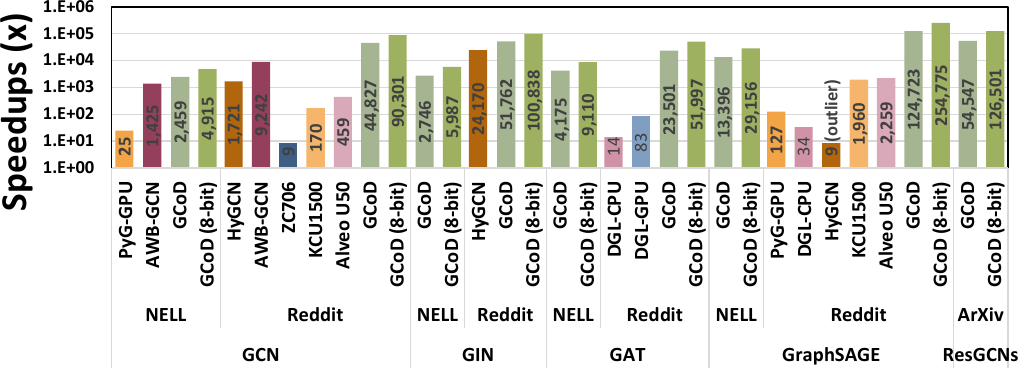}
    \caption{The normalized inference speedups (w.r.t. PyG-CPU) comparisons on \textbf{large} GCN models and graph datasets.}
    \vspace{-1em}
    \label{fig:overall_speedup_large}
\end{figure}

\subsection{Experiment Settings}

\textbf{Models, Datasets, and Training Settings.}
Our evaluation considers \textbf{five GCN algorithms}, including three representative full-batch training GCN algorithms (i.e., GCN \cite{kipf2017semi}, GAT \cite{velivckovic2017graph}, and GIN \cite{xu2018how}),  
one mini-batch training GCN algorithm (i.e., GraphSAGE \cite{hamilton2017inductive}), and one deep ResGCN \cite{li2020deepergcn}, 
and \textbf{six graph datasets}, including three citation graph datasets (Cora, CiteSeer, and Pumbed) \cite{sen2008collective}, the one knowledge graph (NELL) \cite{carlson2010toward},
and the two large scale datasets (Obgn-ArXiv from Open Graph Benchmark (OGB) \cite{hu2020ogb}
and Reddit post dataset \cite{hamilton2017inductive}), respectively.
The specifications of the aforementioned five GCN models are summarized in Tab. \ref{tab:model}, from which we can see that
the adopted GCNs and GINs consist of 16 hidden units for the three citation graphs and 64 hidden units for the NELL and Reddit graphs, following a SOTA GCN accelerator \cite{geng2020awb}; the GAT models consist of 8 hidden units and 8 heads; the GraphSAGE models adopt two layers and the same hidden dimensions as GCNs, with a neighborhood sample size of 25 and 10, respectively, following the basic settings in \cite{PyG, hamilton2017inductive}; and deep ResGCNs adopt 28 layers with 128 hidden units, following the settings in \cite{li2020deepergcn}. 
We train all the above GCN models for 400 epochs using an Adam optimizer \cite{kingma2014adam} with a learning rate of 0.01.
The statistics of these six datasets are summarized in Tab. \ref{tab:dataset}. We follow the same dataset splits as described in \cite{kipf2017semi,hamilton2017inductive,hu2020ogb}.


\begin{table}[t]
  \centering
  \caption{A summary of the adopted graph dataset statistics.}
  \vspace{-0.5em}
  \resizebox{0.48\textwidth}{!}{
    \begin{tabular}{lccccc}
    \hline
    \textbf{Dataset} & \textbf{Nodes} & \textbf{Edges} & \textbf{Features} & \textbf{Classes} & \textbf{Storage} \\
    \hline
    \hline
    Cora & 2,708 & 5,429 & 1,433 & 7   & 15 MB \\
    Citeseer & 3,312 & 4,372 & 3,703 & 6   & 47MB \\
    Pubmed & 19,717 & 44,338 & 500 & 3   & 38MB \\
    NELL & 65,755 & 266,144 & 5,414 & 210 & 1.3GB \\
    Ogbn-ArXiv & 169,343 & 1,166,243 & 128 & 40 & 103MB \\
    Reddit & 232,965 & 114,615,892 & 602 & 41 & 1.8GB\\
    \hline
    \end{tabular}%
    }
  \vspace{-1em}
  \label{tab:dataset}%
\end{table}%

\begin{table}[t]
  \centering
  \caption{A summary of the GCN model specifications.}
  \vspace{-0.5em}
  \resizebox{0.48\textwidth}{!}{
  \begin{tabular}{lcccc}
    \hline
    \textbf{Model} & \textbf{Layers} & \textbf{Hidden Dim.} & \textbf{Aggregation} & \textbf{Others/Details}  \\
    \hline
    \hline
    GCN & 2 & 16/64 & Mean & \multirow{3}[1]{*}{\tabincell{l}{16 for Cora/CiteS./Pub.; \\ 64 for NELL/Reddit}} \\
    GIN & 3 & 16/64 & Add &  \\
    GraphSAGE & 2 & 16/64 & Mean &  \\
    \hline
    GAT & 2 & 8 & Attention & 8 heads \\
    ResGCN & 28 & 128 & Max &  - \\
    \hline
  \end{tabular}%
  }
  \label{tab:model}%
  \vspace{-1em}
\end{table}%

\begin{table}[t]
  \centering
  \caption{System configurations of the baselines and GCoD.}
  \vspace{-0.5em}
  \resizebox{0.48\textwidth}{!}{
    \begin{tabular}{llllrr}
    \specialrule{0.05em}{2pt}{2pt}
     \multicolumn{1}{l}{\textbf{Design/Platform}} & \multicolumn{1}{l}{\textbf{Compute Unit}} & \multicolumn{1}{l}{\textbf{\tabincell{c}{On-Chip Memory}}} & \multicolumn{1}{l}{\textbf{\tabincell{c}{Off-Chip \\ Memory}}} & \multicolumn{1}{l}{\textbf{\tabincell{c}{Die Area \\ (mm$^2$)}}} & \multicolumn{1}{l}{\textbf{\tabincell{c}{Power \\ (W)}}} \\
    \hline
    \specialrule{0.05em}{2pt}{2pt}
    \textbf{\tabincell{l}{PyG/DGL-\\CPU}} &\tabincell{l}{2.5GHz \\ @24 cores} & \tabincell{l}{L1d/L1i: 24 x 32KB \\ L2: 3MB \\ L3: 30MB} & \tabincell{l}{1365.5 GB/s \\ DDR4} & \multicolumn{1}{c}{-} & \multicolumn{1}{c}{150} \\
    \specialrule{0.05em}{2pt}{2pt}
    \textbf{\tabincell{l}{PyG/DGL-\\GPU}} & \tabincell{l}{1.35GHz \\@4352 cores} & \tabincell{l}{L1: 68 x 64KB \\ L2: 5.5MB} & \tabincell{l}{616 GB/s \\ GDDR6} & \multicolumn{1}{c}{\tabincell{c}{754 \\ (12nm)}} & \multicolumn{1}{c}{250} \\
    \specialrule{0.05em}{2pt}{2pt}
    \textbf{HyGCN} & \tabincell{l}{1GHz@32 SIMD\\ \& 8 systolic array} & \tabincell{l}{Input: 128KB; \\ Edge: 2MB; \\ Weight: 2MB; \\ Output: 4MB; \\ Aggregation: 16MB} & \tabincell{l}{256 GB/s \\ HBM$\sim$1.0} & \multicolumn{1}{c}{\tabincell{c}{7.8 \\ (12nm)}} & \multicolumn{1}{c}{6.7} \\
    \specialrule{0.05em}{2pt}{2pt}
    \textbf{AWB-GCN} & \tabincell{l}{330MHz@Intel \\ D5005 FPGA} & \tabincell{l}{4096 PEs \\ 244 Mb Scratchpad} & \tabincell{l}{76.8 GB/s \\ DDR4} & \multicolumn{1}{c}{-} & \multicolumn{1}{c}{215} \\
    \specialrule{0.05em}{2pt}{2pt}
    \textbf{ZC706} & \tabincell{l}{220MHz@900 \\ DSPs} & \tabincell{l}{19.2MB} & \tabincell{l}{12.8GB/s \\ DDR3}   & \multicolumn{1}{c}{-}  &  \\
    \specialrule{0.05em}{2pt}{2pt}
    \textbf{KCU1500} & \tabincell{l}{5520 DSPs} & \tabincell{l}{75.9MB} & \tabincell{l}{76.8GB/s \\ DDR4}   &  \multicolumn{1}{c}{-}  &  \\
    \specialrule{0.05em}{2pt}{2pt}
    \textbf{Alveo U50} & \tabincell{l}{5952 DSPs} & \tabincell{l}{227.3MB} & \tabincell{l}{316GB/s \\ HBM$\sim$2.0}   &  \multicolumn{1}{c}{-} & \multicolumn{1}{c}{50} \\
    \specialrule{0.05em}{2pt}{2pt}
    \textbf{GCoD} & \tabincell{l}{330MHz@\\ VCU128} & \tabincell{l}{\tabincell{l}{4096 PEs\\ 9MB BRAM \\ 33MB URAM}} & \tabincell{l}{460GB/s \\ HBM}   & \multicolumn{1}{c}{\tabincell{c}{-}} & \multicolumn{1}{c}{180}  \\
    \specialrule{0.05em}{2pt}{1pt}
    \end{tabular}%
  }
  
  \label{tab:hardware}%
  \vspace{0.4em}
  \leftline{\small *GCoD (8-bit) uses 10240 PEs as 8-bit saves required bandwidth.}
  \vspace{-1.5em}
\end{table}%

\textbf{Baselines and Evaluation Metrics.}
To benchmark our GCoD with SOTA GCN acceleration works, we consider \textbf{a total of nine baselines}: PyTorch Geometric (PyG) \cite{PyG} and Deep Graph Library (DGL) \cite{wang2019dgl} on a Linux workstation with Intel Xeon E5-2680 v3 CPUs and NVIDIA RTX 8000 GPUs, respectively, and SOTA GCN accelerators HyGCN \cite{yan2020hygcn}, AWB-GCN \cite{geng2020awb}, 
and Deepburning-GL on three FPGA platforms (i.e., ZC706, KCU1500, and Alveo U50) \cite{liang2020deepburning}. The system configurations of the baselines and our  GCoD are summarized in Tab. \ref{tab:hardware}.
We evaluate all above platforms in terms of acceleration latency speedups, energy consumption, and required off-chip memory bandwidth and accesses.
In addition, we compare the achieved accuracy of GCoD algorithm with SOTA GCN compression baselines, including RP \cite{frankle2018the}, SGCN \cite{li2020sgcn}, QAT \cite{fan2020training}, and Degree-Quant \cite{tailor2021degreequant}.

\textbf{Hardware Experiment Setup.}
To evaluate GCoD accelerator, we consider the standard FPGA evaluation and implementation flows in Vivado 2018.3 \cite{vivado}. Specifically, we adopt a Xilinx VCU128 FPGA board \cite{vcu128}, which is equipped with 9024 DSPs, 42MB on-chip memory, and 460 GB/s HBM off-chip memory, where the off-chip memory bandwidth is proportionally distributed across GCoD's sub-accelerators based on their assigned workloads/resources. 
For a fair comparison with AWB-GCN \cite{geng2020awb}, GCoD accelerator is clocked at 330MHz and adopts 4096 PEs with a 32-bit fixed point precision.
In addition, we discuss a GCoD variant that supports the quantized GCNs and termed as GCoD (8-bit). Quantization largely reduces the off-chip memory bandwidth requirement and thus enable GCoD (8-bit) to afford 10240 on-chip PEs ($\approx$ 5200 DSPs).

\begin{figure*}[t]
    \centering
    \includegraphics[width=\linewidth]{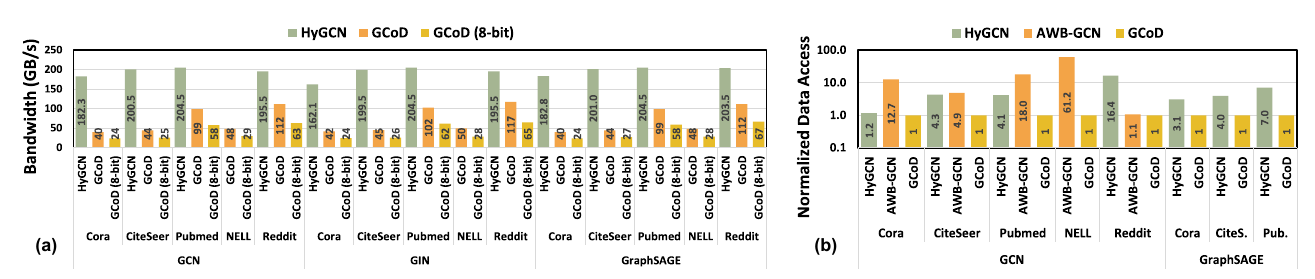}
    \caption{(a) Bandwidth requirement of GCoD and HyGCN and (b) normalized data access of GCoD, HyGCN, and AWB-GCN.
    Note that we record the peak bandwidth in (a), so the relative scale of (a) and (b) are slightly different.}
    \label{fig:bandwidth_comp}
    \vspace{-1em}
\end{figure*}

 \subsection{Overall Performance}
\label{sec:exp:overall}
We first evaluate GCoD against both the general platforms (PyG/DGL-CPU and PyG/DGL-GPU) and SOTA GCN accelerators in terms of speedup, off-chip memory bandwidth requirement, and the number of off-chip memory accesses.

\textbf{GCoD over CPU/GPU Platforms.}
Figs. \ref{fig:overall_speedup} \& \ref{fig:overall_speedup_large} show the overall performance of our GCoD and the baselines. 
We can see that GCoD on-average achieves 15286$\times$, 294$\times$, 1057$\times$, and 460$\times$ speedups over PyG-CPU, PyG-GPU, DGL-CPU, and DGL-GPU, respectively, while GCoD (8-bit) more aggressively achieves 32158$\times$, 607$\times$, 2213$\times$, and 962$\times$ speedups over PyG-CPU, PyG-GPU, DGL-CPU, and DGL-GPU, respectively.
The superior GCoD improvements validate the effectiveness of GCoD's dedicated algorithm and accelerator innovations:
(1) GCoD's split and conquer training algorithm largely alleviates the irregularity of the graph adjacency matrices, leading to more consecutive address accesses of off-chip memory;
and (2) GCoD accelerator enables more balanced workloads and higher utilization for each sub-accelerator, and allows more data reuses and lower on-chip storage demand.

\textbf{GCoD over SOTA GCN Accelerators.}
We further compare GCoD with SOTA GCN accelerators: HyGCN \cite{yan2020hygcn}, AWB-GCN \cite{geng2020awb}, and Deepburning-GL on three FPGA platforms (ZC706, KCU1500, and Alveo U50) \cite{liang2020deepburning}. We follow most of these baselines to report the relative speedups over PyG-CPU on an Intel Xeon E5-2680 v3 CPU, and analyze the achieved improvements, as elaborated below.


\begin{table}[t]
  \centering
  \caption{\hr{Speedup breakdown of GCoD accelerator w/ or w/o sparsification (SP.) and quantization (Quant.).}}
  \vspace{-0.5em}
  \resizebox{0.48\textwidth}{!}{
    \begin{tabular}{lccccc}
    \hline
    \multicolumn{1}{c}{\multirow{2}[4]{*}{\textbf{Methods}}} & \multicolumn{5}{c}{\textbf{Speedups over PyG-CPU}}  \\
    \cline{2-6}  & \textbf{Cora} & \textbf{CiteSeer} & \textbf{Pubmed} & \textbf{NELL} & \textbf{Reddit}  \\
    \hline
    \hline
    AWB-GCN & 1063$\times$ & 913$\times$ & 466$\times$ & 1425$\times$ & 9242$\times$ \\
    \textbf{GCoD Accele.} &1824$\times$   & 1692$\times$   & 901$\times$   &  2294$\times$  & 39881$\times$  \\
    \textbf{GCoD Accele. w/ SP.} & 2031$\times$  & 1763$\times$  & 970$\times$  & 2459$\times$  & 44827$\times$  \\
    \textbf{GCoD Accele. w/ SP. \& Quant.} & 4373$\times$  & 3459$\times$  & 1931$\times$  & 4915$\times$  & 90301$\times$  \\
    \hline
    \end{tabular}%
  }
  \label{tab:breakdown}%
  \vspace{-1.em}
\end{table}%

(1) Speedup. 
As shown in Fig. \ref{fig:overall_speedup} \& \ref{fig:overall_speedup_large}, 
GCoD on-average achieves 7.8$\times$, 2.5$\times$, 2532$\times$, 165$\times$, and 115$\times$ speedups over HyGCN (without considering HyGCN's outlier when evaluated GraphSAGE), AWB-GCN, and Deepburning-GL on three FPGA platforms (ZC706, KCU1500, and Alveo U50), respectively.
GCoD benefits are attributable to the dedicated algorithm and accelerator co-design. Specifically, HyGCN adopts coarse-grained block-wise scheduling while GCoD adopts fine-grained, adaptive row/column-wise pipeline; AWB-GCN realizes workload balance via on-the-fly auto-tuning while GCoD leverages a split and conquer algorithm to achieve naturally balanced workload;
\hr{We further provide the improvement breakdown in Tab. \ref{tab:breakdown}, from which we can see that the improvement is mostly attributed to GCoD's two-pronged accelerator that leads to on-average 2.29$\times$ speedup over AWB-GCN \cite{geng2020awb}, while sparsification further provides 1.09$\times$ speedups. Meanwhile,
the GCoD (8-bit) variant offers an additional 2.02$\times$ on-average speedup.}

(2) Memory Bandwidth Consumption and Accesses. Fig. \ref{fig:bandwidth_comp} (a) shows the evaluation results in terms of off-chip memory bandwidth consumption. We can see that GCoD and GCoD (8-bit) only require on-average 48\% and 26\% off-chip memory bandwidth as compared to HyGCN, respectively. The high bandwidth of HyGCN is attributed to the required high-degree parallelism, whereas GCoD accelerator's fine-grained pipelines enable more frequent data reuses, largely alleviating the off-chip bandwidth requirement. Fig. \ref{fig:bandwidth_comp} (b) reports 
\hr{the measured off-chip memory accesses comparison for processing GCNs with GCoD, HyGCN, and AWB-GCN.
Note that we count the number of off-chip accesses assuming that the input features and adjacency matrices are stored in the off-chip memory before processing. In practice, these matrices can be partially or entirely stored on-chip to reduce the data accesses and bandwidth requirements \cite{geng2020awb}.}

\begin{table}[t]
  \centering
  \caption{\hr{Comparison between GCoD with SOTA GCN compression methods, including Random Pruning (RP) \cite{frankle2018the}, SGCN \cite{li2020sgcn}, QAT \cite{fan2020training}, and Degree-Quant \cite{tailor2021degreequant}.}}
  \vspace{-0.5em}   
  \resizebox{0.48\textwidth}{!}{
\begin{tabular}{clccccc}
\specialrule{0.05em}{1pt}{1pt}
\multirow{2}[4]{*}{\textbf{Models}} & \multicolumn{1}{c}{\multirow{2}[4]{*}{\textbf{Methods}}} & \multicolumn{5}{c}{\textbf{Accuracy (\%)}} \\
\cline{3-7} \specialrule{0em}{1pt}{1pt}  &   & \multicolumn{1}{c}{\textbf{Cora}} & \multicolumn{1}{c}{\textbf{CiteSeer}} & \multicolumn{1}{c}{\textbf{Pubmed}} & \multicolumn{1}{c}{\textbf{NELL}} & \multicolumn{1}{c}{\textbf{Reddit}}\\
\specialrule{0.05em}{1pt}{1pt}
\specialrule{0.05em}{1pt}{1pt}
\multirow{8}[4]{*}{GCN} & Vanilla & 81.1$\pm$1.2 & 70.2$\pm$0.8 & 79.1$\pm$0.6 & 65.6$\pm$0.7 & 92.2$\pm$1.1 \\
\cdashline{2-7}
  & RP & 79.6$\pm$0.8  & 70.4$\pm$0.5  & 78.4$\pm$0.6 & 63.5$\pm$2.3 & 91.2$\pm$2.2 \\
  & SGCN & 80.2$\pm$0.7  & 70.4$\pm$0.7  & 79.1$\pm$0.1 & 64.2$\pm$1.2 & 91.3$\pm$1.3  \\
  & QAT & 81.0$\pm$0.7  & 71.3$\pm$1.0  & 79.0$\pm$0.2 & 65.1$\pm$1.4 & 92.4$\pm$0.9 \\
  & Degree-Quant & 81.7$\pm$0.7  & 71.0$\pm$0.9  & 79.1$\pm$0.1 & 65.2$\pm$0.8 & 92.6$\pm$1.5 \\
  & \textbf{\hr{GCoD}} & \textbf{81.9$\pm$0.8}  & \textbf{71.7$\pm$0.5}  & 79.5$\pm$0.3 & \textbf{66.3$\pm$0.5} & \textbf{93.4$\pm$0.9}  \\
  & \textbf{GCoD (8-bit)} & 81.0$\pm$0.9  & 70.6$\pm$0.3  & \textbf{79.5$\pm$0.2} & 66.0$\pm$0.3 & 93.2$\pm$1.3  \\
\cline{2-7} 
\specialrule{0em}{1pt}{1pt} & \textbf{GCoD Improv.} &  \multicolumn{5}{c}{\textbf{$\uparrow$0.2\% $\sim$ $\uparrow$2.8\%}} \\
\specialrule{0.05em}{1pt}{1pt}
\multirow{8}[4]{*}{GAT} & Vanilla & 83.1$\pm$0.4  & 72.2$\pm$0.7  & 78.8$\pm$0.3 & 66.6$\pm$0.3 &  94.2$\pm$0.3 \\
\cdashline{2-7}
  & RP & 80.9$\pm$0.6  & 69.8$\pm$0.8  & 78.2$\pm$0.1  & 64.5$\pm$1.2 & 93.1$\pm$1.2 \\
  & SGCN & 81.9$\pm$0.3  & 71.9$\pm$0.2  & 78.4$\pm$0.1  & 64.9$\pm$1.0 & 93.4$\pm$0.9  \\
  & QAT & 81.9$\pm$0.7  & 71.2$\pm$1.0  & 78.3$\pm$0.5 & 65.1$\pm$0.8 & 93.8$\pm$0.5 \\
  & Degree-Quant & 82.7$\pm$0.7  & 71.6$\pm$1.0  & 78.6$\pm$0.3  &  65.9$\pm$0.6 & 94.0$\pm$0.7 \\
  & \textbf{\hr{GCoD}} & \textbf{83.2$\pm$0.3}  & \textbf{72.2$\pm$0.4}  & \textbf{79.0$\pm$0.1} & \textbf{66.7$\pm$0.4} & 94.5$\pm$0.2 \\
  & \textbf{GCoD (8-bit)} & 82.6$\pm$0.2  & 71.8$\pm$0.1  & 78.8$\pm$0.2 & 66.5$\pm$0.2 & \textbf{94.5$\pm$0.4} \\
\cline{2-7} 
\specialrule{0em}{1pt}{1pt} & \textbf{GCoD Improv.} &  \multicolumn{5}{c}{\textbf{$\uparrow$0.1\% $\sim$ $\uparrow$2.2\%}} \\
\specialrule{0.05em}{1pt}{1pt}
\multirow{8}[4]{*}{GIN} & Vanilla & 78.6$\pm$0.9  & 67.5$\pm$1.5  & 78.5$\pm$0.2  & 65.2$\pm$0.2 & 92.8$\pm$2.2\\
\cdashline{2-7}
  & RP & 74.6$\pm$0.4  & 64.5$\pm$0.5  & 76.9$\pm$0.6  & 64.2$\pm$0.5 & 92.0$\pm$0.5 \\
  & SGCN & 78.0$\pm$0.1  & 67.0$\pm$0.1  & 77.2$\pm$1.1 & 64.8$\pm$0.4 & 92.3$\pm$0.9 \\
  & QAT & 75.6$\pm$1.2  & 63.0$\pm$2.6  & 77.5$\pm$0.2   & 64.7$\pm$0.3 & 92.9$\pm$0.4\\
  & Degree-Quant & 78.7$\pm$1.4  & 67.5$\pm$1.4  & 78.1$\pm$0.5  & 65.2$\pm$0.3 & 93.1$\pm$0.6\\
  & \textbf{\hr{GCoD}} & \textbf{78.9$\pm$0.5}  & 68.6$\pm$0.8  & \textbf{78.5$\pm$0.3}  & \textbf{65.8$\pm$0.2} &  93.3$\pm$0.9  \\
  & \textbf{GCoD (8-bit)} & 78.4$\pm$0.2  & \textbf{68.7$\pm$1.3}  & 78.3$\pm$0.2  & 65.6$\pm$0.4 &  \textbf{93.3$\pm$1.1}  \\
\cline{2-7}  
\specialrule{0em}{1pt}{1pt} & \textbf{GCoD Improv.} &  \multicolumn{5}{c}{\textbf{$\uparrow$0.3\% $\sim$ $\uparrow$4.2\%}} \\
\specialrule{0.05em}{1pt}{1pt}
\multirow{5}[4]{*}{GraphSAGE} & Vanilla & 81.2$\pm$0.2  & 71.1$\pm$0.3  & 78.7$\pm$0.2  &  66.2$\pm$0.4  & 93.8$\pm$2.9 \\
\cdashline{2-7}
  & RP & 77.7$\pm$0.7  & 66.1$\pm$2.2  & 76.0$\pm$0.3  & 65.5$\pm$0.4 & 90.7$\pm$1.8 \\
  & SGCN & 79.2$\pm$0.5  & 70.9$\pm$0.3  & 78.5$\pm$0.1  & 66.1$\pm$0.3 & 93.9$\pm$0.9 \\
  & \textbf{\hr{GCoD}} & \textbf{81.4$\pm$0.6}  & 71.3$\pm$0.3  & \textbf{79.0$\pm$0.4}  & \textbf{66.7$\pm$0.6} &  \textbf{94.5$\pm$1.2} \\
  & \textbf{GCoD (8-bit)} & 80.8$\pm$0.4  & \textbf{71.3$\pm$0.1}  & 78.8$\pm$0.1  & 66.4$\pm$0.8 &  94.3$\pm$1.5 \\
\cline{2-7}
\specialrule{0em}{1pt}{1pt} & \textbf{GCoD Improv.} &  \multicolumn{5}{c}{\textbf{$\uparrow$0.2\% $\sim$ $\uparrow$3.8\%}} \\
\specialrule{0.05em}{1pt}{1pt}
\end{tabular}%
    }
    \vspace{-1.em}
  \label{tab:GCoD_alg}%
\end{table}%


\subsection{Evaluation of the GCoD Algorithm}
\label{sec:exp:algorithm}

\textbf{GCoD over SOTA Compression Baselines.}
Table \ref{tab:GCoD_alg} compares the accuracy of GCoD with SOTA compression methods to evaluate the effectiveness of GCoD algorithm.
We can see that GCoD consistently achieves a comparable or even better accuracy \hr{($\uparrow$0.2\% $\sim$ $\uparrow$4.2\%)} over the vanilla GCNs and all compression baselines, while offering a 5\% $\sim$ 15\% structural sparsity ratio (and thus more balanced workload).

\textbf{Ablation Studies of the Design Hyper-Parameters.}
Our GCoD algorithm has two hyper-parameters: the total number of classes $C$ (i.e., sub-accelerators) and subgraphs $S$.
To validate sensitivity of GCoD benefits, we measure the speedups and off-chip memory bandwidth requirements across a wide range of the design hyper-parameters $C \in \{1, 2, 3, 4\}$ and $S \in \{8, 12, 16, 20\}$, and find that
GCoD consistently achieves 1.8$\times$ $\sim$ 2.8$\times$ speedups over AWB-GCN and reduces the off-chip memory bandwidth by 26\% $\sim$ 53\%, validating GCoD's general effectivenss and robustness.

\begin{figure}[t]
    \centering
    \includegraphics[width=\linewidth]{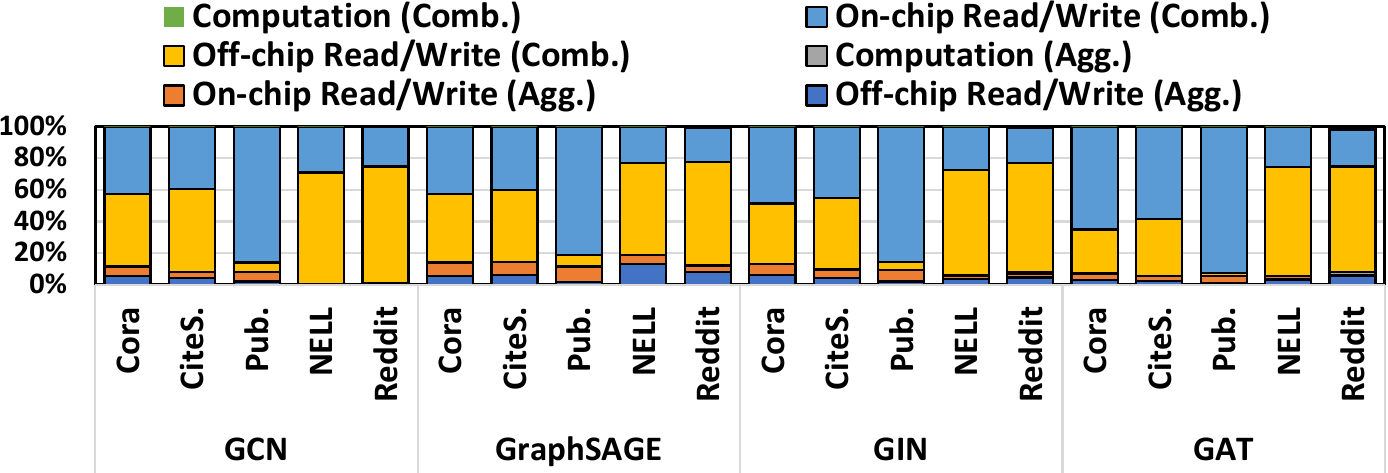}
        \caption{The energy breakdown of the GCoD framework when evaluated on the four GCN models and five graph datasets.}
    \label{fig:overall_energy}
    \vspace{-1.5em}
\end{figure}

\subsection{Evaluation of the GCoD Accelerator}
\label{sec:exp:hardware}

\textbf{Energy Breakdown.}
Fig. \ref{fig:overall_energy} shows GCoD accelerator's energy breakdown in terms of computations and off-chip memory accesses in both the combination and aggregation phases.
We can see that 
(1) the combination phase consumes most of the energy than the aggregation phase, thanks to GCoD acceleration (vs. PyG-CPU on which aggregation occupies 80\% $\sim$ 99\% \cite{yan2020hygcn}), indicating the effectiveness of GCoD in alleviating the performance bottleneck due to the aggregation phase, and 
(2) the energy cost of accessing HBM remains reasonable
as graph size increases, validating GCoD's scalability.

\textbf{Resource-aware vs. Efficiency-aware Pipeline.}
As discussed in Sec. \ref{sec:subaccelerator_arch}, GCoD adopts efficiency-aware pipeline for small/medium graphs to achieve more data reuses at a cost of storing aggregation outputs on-chip. When processing large graphs, e.g., Reddit the outputs of which require 36MB storage and cannot be fully stored on-chip, GCoD uses resource-aware pipeline for better balancing the data reuses and on-chip storage requirement. The relatively more off-chip memory accesses when processing Reddit (see Fig. \ref{fig:bandwidth_comp} (b)) is resulting from the less data resues for reduced on-chip storage. 

\section{Conclusion}
We propose, develop, and validate GCoD, an algorithm and accelerator co-design framework. 
On the algorithm level, 
GCoD integrates a split and conquer GCN training strategy to polarize the graphs to be either denser or sparser in local neighborhoods without compromising the model accuracy, resulting in graph adjacency matrices that have merely two levels of balanced workload and thus enjoy largely enhanced regularity.
On the hardware level, GCoD integrates a dedicated two-pronged accelerator with a dedicated engine to process each of the aforementioned workloads, further boosting the overall utilization and acceleration efficiency.
Extensive experiments and ablation studies validate the advantages of GCoD.
\section*{Acknowledgment}

We would like to acknowledge the funding support from the NSF RTML program (Award number: 1937592) and the NSF NeTS program (Award number: 1801865) for this project.
The authors also thank our colleague Mr. Cheng Wan at Rice University for his help and discussion in the graph reordering algorithm.

\bibliographystyle{IEEEtranS}
\bibliography{refs}




\end{document}